# Enhanced Quantum Effects in an Ultra-Small Coulomb Blockaded Device Operating at Room-Temperature

S. J. Shin,[1] C. S. Jeong,[1] B. J. Park,[1] T. K. Yoon,[1] J. J. Lee,[1] S. J. Kim,[1] J. B. Choi,[1*] Y. Takahashi,[2] and D. G. Hasko[3]

[1]*Department of Physics & Research Institute for Nano Science & Technology, Chungbuk National University, Cheongju 361-763, Republic of Korea.*
[2]*The Graduate School of Information Science and Technology, Hokkaido University, Sapporo 060-0814, Japan.*
[3]*Cavendish Laboratory, Univ. of Cambridge, Cambridge CB3 0HE, United Kingdom.*
* e-mail: jungchoi@chungbuk.ac.kr

An ultra-small Coulomb blockade device can be regarded as a mesoscopic artificial atom system and provides a rich experimental environment for studying quantum transport phenomena[1]. Previously, these quantum effects have been investigated using relatively large devices at ultra-low temperatures, where they give rise to a fine additional structure on the Coulomb oscillations[2-13]. Here, we report transport measurements carried out on a sub-2nm single-electron device; this size is sufficiently small that Coulomb blockade, and other quantum effects, persist up to room temperature (RT). These devices were made by scaling the size of a FinFET structure down to an ultimate limiting form, resulting in the reliable formation of a sub-2nm silicon Coulomb island. Four clear Coulomb diamonds can be observed at RT and the 2$^{nd}$ Coulomb diamond is unusually large, due to quantum confinement. The observed characteristics are successfully modeled on the basis of a very low electron number on the island, combined with Pauli spin exclusion. These effects offer additional functionality for future RT-operating single-electron device applications.

The availability of an ultra-small Coulomb island could provide new operating regimes for the single-electron transistor (SET); where Coulomb blockade and quantum confinement are two distinct effects, competing to control electron transport through a device. As the island size is reduced below 2nm, the energy scale for quantum confinement (QC) becomes dominant over both the Coulomb charging and thermal energies at room temperature (RT). Under these conditions, QC may be used as an additional state variable to control electron transport. Moreover, the very small number of electrons on the island ensures that Pauli spin exclusion also strongly influences the electron transport characteristics. Here, quantum effects provide an additional resource to be exploited in nanoscaled SETs, in contrast to conventional CMOS, where quantum effects usually contribute only problems, such as noise and unwanted threshold voltage shifts. Significant earlier work has been aimed at implementing a room-temperature operating SET,[14-22] using various schemes, device structures and fabrication processes,[23-32] but reliable processes for the controlled fabrication of ultra-small size Coulomb islands of less than 5nm have not been firmly established yet. Most of the gate-dependent drain currents exhibited only a single NDR-like peak, or somewhat wave-like shapes of Coulomb oscillations with very poor peak-to-valley current ratios (PVCRs) <<1, leading to very weak quantum effects and limiting practical device applications at room-temperature.

We have modified a state-of-the-art silicon FinFET structure[33] to scale the device to an ultimate form, by using deep-trench and pattern-dependent oxidation, which can be used to form an SET with a Coulomb island size of less than 5nm. By wrapping a fin-gate almost completely around this island, good control of the local electron potential is maintained. Figure 1a shows a schematic 3-D layout of the device; the fabrication process is summarized in the Methods section and further details are given in the Supplementary Information (SI) section. A cross-sectional schematic view along cutline a-b is seen in Fig. 1b; note how the top-Si nanowire, exposed by the nano-gap between the source and the drain, is further etched down to <5nm in depth, by dry etching and gate oxidation. This key process step enables an island to be formed with nearly identical tunnel barriers in a self-aligned manner[34]. Figure 1c shows a TEM image of the top-Si nanowire active channel, the poly-Si fin-gate and the TEOS spacer profiles along the cutline a-b. The fin-gate above the center of the etched region of the active channel is about 10-nm wide and wraps most of the way around the active channel. This gate is important for realizing a large PVCR in the Coulomb oscillations at RT. Figures 1d and 1e show TEM cross-sectional images of the etched top-Si nanowire along the cutline c-d after oxidation at 900℃ for 50min and 40min and the island size has reduced to ~2-nm (SET_A) and ~5-nm (SET_B) respectively. Good control over the island size was achieved through this oxidation process (see Fig. S3 in the SI).

Figure 2a shows Coulomb oscillations from SET_A measured at room temperature; these characteristics show high, and nearly symmetric, PVCRs, indicating that the tunnel barriers are nearly identical. The temperature dependence of these characteristics is shown in Fig. 2b; as the temperature is reduced, the Coulomb peak positions barely change, but the PVCRs increase. As seen in Fig. 2a & 2b, the magnitudes of the first two current peaks are observed to be relatively weak compared to the other 3$^{rd}$ and 4$^{th}$ Coulomb peaks, indicating that the gate coupling becomes considerably weaker at high Vg. This could be attributed to the specific structure of the gate and Coulomb channel in our device. As seen in the Fig. 1c) showing a cross-sectional TEM images along the cutlines a-b, the fin-gate is actually T-shaped. The lower part of the fin-gate located right above the center of the trenched region of the active channel is about 10-nm wide and wraps most of the way around the active channel, which results in a large peak-to-valley PVCR in the Coulomb oscillations at RT. However, the other upper parts of the fin-gate are located very far from the active channel and cover large areas of the 2DEG source and drain regions, rendering these ineffective at low values of Vg. However, as Vg is increased, it becomes effective for gate voltages greater than ~8V, so affecting the 2DEG potential of the source, drain and also the potential barrier, leading to reduced coupling with the channel. Moreover, in our device, the confinement along the axis of the Coulomb channel is somewhat weaker than that in the perpendicular directions (Si/SiO2) and the confinement length scale must depend on the gate voltage for this direction. The density of states parallel to the channel axis must increase rapidly with gate voltage, leading to a rapidly increasing tunnel barrier transmission at high Vg.

A room temperature charge stability plot is shown in Fig. 2c, where three and a half Coulomb diamonds are clearly seen; each diamond corresponds to a stable charge configuration on the Coulomb island. The number of occupied electrons (N) for each diamond is estimated as follows. As seen in Fig. 2a, the first current peak appears at $V_G$~3V. Its magnitude is relatively weak compared to other Coulomb peaks, but is seen for all finite drain bias voltages. The small size of this peak could be due to the effect of the fin-gate bias voltage on the transmission properties of the tunnel barriers. As the gate voltage is increased, there is a corresponding increase in the transmission of the tunnel barrier, so that current suppression between the Coulomb peaks becomes ineffective for gate voltages greater than ~8V in SET_A at room temperature. Nevertheless, extrapolating the expected transmission at the location where the next peak in the series below N=1 would occur suggests that such a peak would still be detectable. The lack of any feature in this area indicates that the island is unpopulated by electrons for $V_G$≤3V. This view is supported by the Coulomb diamond data of Fig. 2c, where as the gate voltage is made less positive, the slope of each Coulomb diamond steeply increases, and the Coulomb diamond (for $V_G$<3V) does not close. This important assumption on the dot occupancy should be further discussed with reference to our adapting 3D box model for a Si dot regarding the electron filling and level spacing, which will be detailed later. In the 3D box model, the ground state is non-degenerate and the 1$^{st}$ level spacing ΔE (>> kbT) agrees to the measured shift of the 3$^{rd}$ Coulomb peak associated with quantum confinement. This allows the first two electrons to fill the non-degenerate ground state (associated with the first two diamonds in Fig. 1c) in a spin-up-spin-down sequence. The magnetic field dependence of these peaks may help to resolve this matter. The Zeeman splitting in Si, however, is extremely small, $g\mu_B B = 1.16 \times 10^{-4}$(eV/T)B, so that ultra-low temperatures (a few mK) and high magnetic fields (of order 10Tesla) would be needed to resolve the relevant splitting, which is beyond our research scope.

It is noted that the 2$^{nd}$ Coulomb diamond (corresponding to N=2) has an unusually large diagonal size of $\Delta V_G \sim 3.8V$. In contrast with the sizes of the other diamonds that are determined mainly by Coulomb charging, the abnormal size of the 2$^{nd}$ diamond is due to a shift of the 3$^{rd}$ Coulomb peak, caused by an additional QC effect. In order to understand these characteristics, we first estimate the size of the island and calculate the Coulomb charging and QC energies. Values for the gate and junction capacitances can be obtained from the Coulomb peak spacing $\Delta V_G$ and the slopes of the 1$^{st}$ Coulomb diamond (which are mainly determined by the Coulomb charging energy and so correspond closely to the orthodox capacitive model)[22]. We find that $C_G \sim 0.094aF$, $C_S \sim 0.16aF$ and $C_D \sim 0.17aF$, yielding the total capacitance $C_\Sigma \sim 0.42aF$, which corresponds to a 1.58nm diameter spherical silicon dot, in good agreement with the TEM image in Fig. 1d. The charging energy of a dot of this size is $e^2/C_\Sigma \sim 0.377eV$, which is more than one order magnitude larger than the thermal energy at room temperature. The level spacing $\Delta E^{QC}$ due to QC for a silicon dot is estimated based on the effective mass approach. Because crystalline silicon is anisotropic with two effective masses of $m_\perp^*$ and $m_\parallel^*$, the Coulomb island is better be modeled by a cube with an effective side length of $L_{eff}=(4\pi/3)^{1/3}d$, where d is the diameter of the dot assumed to be spherical. For a silicon island embedded in $SiO_2$, the quantized level spacing

$\Delta E^{QC} = E(1,1,2) - E(1,1,1)$, where
$$E(n_1,n_2,n_3) = \frac{\pi^2 \hbar^2}{2L_{eff}^2}\left(\frac{n_1^2}{m_\parallel^*} + \frac{n_2^2}{m_\parallel^*} + \frac{n_3^2}{m_\perp^*}\right)$$
, where $m_\perp^* = 0.916 m_0$ and $m_\parallel^* = 0.19 m_0$. Figure 3a shows calculations of the dependence of the Coulomb charging energy $e^2/C_\Sigma$ and the level spacing $\Delta E^{QC}$ due to the QC effect on the island size. For relatively large dots ($L_{eff} > 10nm$), $\Delta E^{QC}$ is small compared with $e^2/C_\Sigma$, and QC causes only a slight modulation of the conventional Coulomb oscillations. However, since the magnitude of the QC effect depends on the inverse square of $L_{eff}$, this effect becomes dominant at the few nm size scale. SET_A corresponds to an $L_{eff}=1.58nm$, so that $\Delta E^{QC} \sim 0.493eV$, which is significantly larger than the Coulomb energy $e^2/C_\Sigma \sim 0.377eV$ for this device.

We now use these results to explain how the abnormal size of the 2$^{nd}$ Coulomb diamond (N=2) arises due to the shift of the 3$^{rd}$ Coulomb peak caused by QC and Pauli spin exclusion. Figure 3b shows schematic energy level diagrams and possible transport mechanisms; each diagram corresponds to one of the gate voltage regions seen in Fig. 2c. It is assumed that the island is empty (N=0) for the gate voltage region $V_G < 2.9V$ and that only one electron is present (N=1) for the gate voltage region $2.9V < V_G < 4.6V$. For a 2$^{nd}$ electron to enter the island, only the Coulomb charging energy $e^2/C_\Sigma \sim 0.377eV$ is needed (as depicted in (i) and (ii)) because the ground state can be occupied by up to two electrons with opposite spin. This corresponds to a gate voltage interval of $\Delta V_G = e/\alpha_G C_\Sigma \sim 1.70V$ between the 1$^{st}$ and 2$^{nd}$ Coulomb peaks, where $\alpha_G$ is a coupling factor defined by $\alpha_G = C_G/C_\Sigma \sim 0.187$. However, the 3$^{rd}$ electron is blocked from entering the ground state defined by the Coulomb charging, due to the need to satisfy the Pauli spin exclusion principle. As the island now has one spin up and one spin down electron in the ground state, the next electron must tunnel onto the 1$^{st}$ excited state. This tunnel event needs an energy of $\Delta E^{QC}$ in addition to the Coulomb energy $e^2/C_\Sigma$, as seen in Fig. 3b(iii); this illustrates the transport behavior expected around the 3$^{rd}$ peak, where the island occupancy oscillates between N=2 and N=3. Taking $\Delta E^{QC}$~0.493eV, the additional gate voltage shift for the 3$^{rd}$ Coulomb peak, due to $\Delta E^{QC}$, is estimated to be $\Delta V_G^{QC}=\Delta E^{QC}/e\alpha_G$~2.24V. Hence, the total shift of the gate voltage of the 3$^{rd}$ peak is expected to be $\Delta V_G=e/\alpha_G C_\Sigma +\Delta V_G^{QC}$~3.94V, which is in good agreement with the measured value of ~3.84V. Note that $\Delta V_G$ between the 3$^{rd}$ and 4$^{th}$ Coulomb peaks is restored to that corresponding to the Coulomb charging energy only. This is consistent with the Pauli spin exclusion principle as the 4$^{th}$ electron is able to tunnel onto the other spin state available in the 1$^{st}$ excited state, and needs only the Coulomb charging energy, as depicted in Fig. 3b(iv). It should be noted that the charging energy could be slightly increased due to a small decrease in the dot capacitance as the fin-gate voltage is increased up to 10V. This is closely related to the gate coupling that becomes weaker at high Vg, and could be attributed to the specific structure of the fin-gate in our device. Unlike the lower part of the fin-gate, the upper parts of the fin-gate are located very far from the active channel and cover large areas of 2DEG source and drain regions. This upper part of the fin-gate, is ineffective at low Vg, and becomes effective only when gate voltages greater than ~8V are used, leading to a parasitic capacitance $C_P$ to the 2DEG source and drain. Since this capacitance is connected to the island capacitance $C_\Sigma$ in series and $C_p >> C_\Sigma$, the total effective capacitance $C_\Sigma^{eff}$ (=1/[1/$C_\Sigma$ +1/$C_p$] ~$C_\Sigma[1-C_\Sigma/C_p]$) becomes subject to a small decrease for high Vg, leading to the observed about 13% increase in the charging energy in the N=3 diamond.

This interpretation of the enhanced quantum effect is confirmed by measurements on another sample (SET_B), which showed an almost identical additional $\Delta V_G^{QC}$ at temperatures of ≤200K. The island size for SET_B is ~5 nm in diameter (see the image in Fig. 1e), so that in this case the QC effect is smaller than the Coulomb charging energy.

Figure 4a shows the temperature dependence of SET_B; at 300K, only one Coulomb peak is resolved, indicating that the Coulomb charging energy is not sufficient to overcome thermal effects. However, at temperatures below 200K well-defined Coulomb oscillations are displayed and with increasing PVCR as the temperature is reduced. In this case, the $V_G$ interval between the 2nd and 3rd Coulomb peaks is $\Delta V_G$ ~1.30V, which is ~30% larger than the other two $V_G$ spacings. Following the previous analysis for SET_A, this additional part of $\Delta V_G$ of ~0.30V is attributed to the QC effect, while the main part of $\Delta V_G$ (usually ~1.0V) is attributed to the Coulomb charging energy. Figure 4b shows the drain bias dependence of SET_B measured at 200K. Following the same arguments used with SET_A, the additional gate voltage shift of ~0.30V, observed for N=2 region, results from the $\Delta E^{QC}$ corresponding to the addition energy for the 3rd electron to tunnel onto the 1$^{st}$ excited state of the dot. We estimate $C_\Sigma$ ~1.4aF for SET_B, which corresponds to $L_{eff}$~5.2nm, so that $\Delta E^{QC}$~0.045eV and the Coulomb charging energy is $C_\Sigma$~0.114eV. Hence, the additional gate voltage shift of the 3$^{rd}$ Coulomb peak is estimated to be $\Delta V_G^{QC}$~0.39V, which is in fair agreement with the measured value of ~0.30V. The resulting values for $\Delta E^{QC}$ and $e^2/C_\Sigma$ for SET_B are illustrated in Fig. 3a, which can be compared with those for SET_A. Note that the QC effect is dramatically reduced for SET_B compared with SET_A due to the difference in dot size; the QC effect is expected to be negligible for sizes >~10nm.

The validity of the effective mass approach for an ultra-small dot of <5nm should be addressed with reference to our adapting 3D box model regarding the electron filling and level spacing. The 3D quantum box must contain enough silicon atoms so that the effective mass approximation and the static value for permittivity are applicable. Theses conditions have been proved to be generally accepted for silicon spheres with radius in excess of 1nm.[35-37] Horiguchi[36] have shown that, with appropriate boundary conditions, effective mass theory gives very similar results to a first-principles calculation for silicon nanowires down to at least 1nm. Extensive work applying effective mass theory to Si-based 2D or 1D systems has shown that the sixfold valley degeneracy is lifted. For an instance, Zheng et al.[37] reported that for 1.2nm thick Si-wire the sixfold valley degeneracy is lifted to form twofold and fourfold degenerate valleys with a splitting of 285meV. For a 3D quantum box, the degenerate valleys are expected to be further split because splitting occurs due to the anisotropy in the effective masses and in the quantum confinements. However, the lifting of degeneracy is a very challenging theoretical problem because its magnitude strongly depends on the details of the confinement characteristics; and on the finiteness of potential barriers and on the stresses, which are highly anisotropic.[38] Our adaptable 3D box model, with anisotropic effective masses, is clearly the simplest possible, but much more appropriate than that of the conventional isotropic hard sphere. In this 3D box model, the ground state E(111) is non-degenerate and the level splitting of E(112)-E(111) agrees with the measured shift of the 3$^{rd}$ peak associated with the quantum confinement. This allows the first few electrons to fill the low energy levels in a spin-up-spin-down sequence. Despite considerable merit in simplicity, however, a more detailed energy spectrum, the level degeneracy and the density of states for this 3D box model would be complicated in reality. In our ultra-small Coulomb island, the highly anisotropic confinement and strain effects due to the pattern-dependent oxidation should be reflected in this context.

In summary, we report on the CMOS-compatible fabrication and behavior of ultra-small silicon SET devices. A modified Fin-FET process was used to scale the size of the Coulomb island down to the sub-2nm regime. At this size scale, both the Coulomb charging and quantum confinement energies are larger than the thermal energy at room temperature. Pauli spin exclusion is shown to strongly influence island filling and offers a new operating regime for these ultra-small Si devices. Such devices allow RT-operating ultra-low power IC and bio-inspired neuromorphic system applications, or simple spin manipulation and readout strategies for quantum computing applications.

## METHODS

Devices were fabricated using a 50-nm-thick undoped silicon-on-insulator (SOI) substrate. First, a 20-nm-wide nanowire, connecting the S/D electrodes, was formed using electron-beam lithography (EBL) with PMMA resist and subsequent RIE etching using $SF_6/CF_4/O_2$. After deposition of a 100nm-thick TEOS layer, an 80-nm-wide trench was etched across the nanowire using EBL with ZEP520A resist and subsequent RIE etching using $CHF_3/O_2$. The exposed silicon area under the trench was etched by a further 30nm in depth by dry etching using $SF_6/CF_4/O_2$, and followed by gate oxidation at 900℃ for 50min (SET_A) and 40min (SET_B). See Fig. 1d & 1e for TEM images of the cross-sectional view of the etched wires. Next, a 20 nm thick TEOS layer was deposited into the trench and etched to form spacers on the sidewalls of the source and drain electrodes. Doped poly-silicon was then deposited into the trench to form a self-aligned fin-gate very close to the Coulomb island. The fabrication was completed with conventional S/D ion implants, annealing and contact processes.

**Acknowledgements** We acknowledge National NanoFab (NNF) for partial technical support. This work was supported by the National Research Foundation (NRF) through the Frontier 21 National Program for Tera-level NanoDevices (TND) and Global Partnership Research Program (GPP) with University of Cambridge.

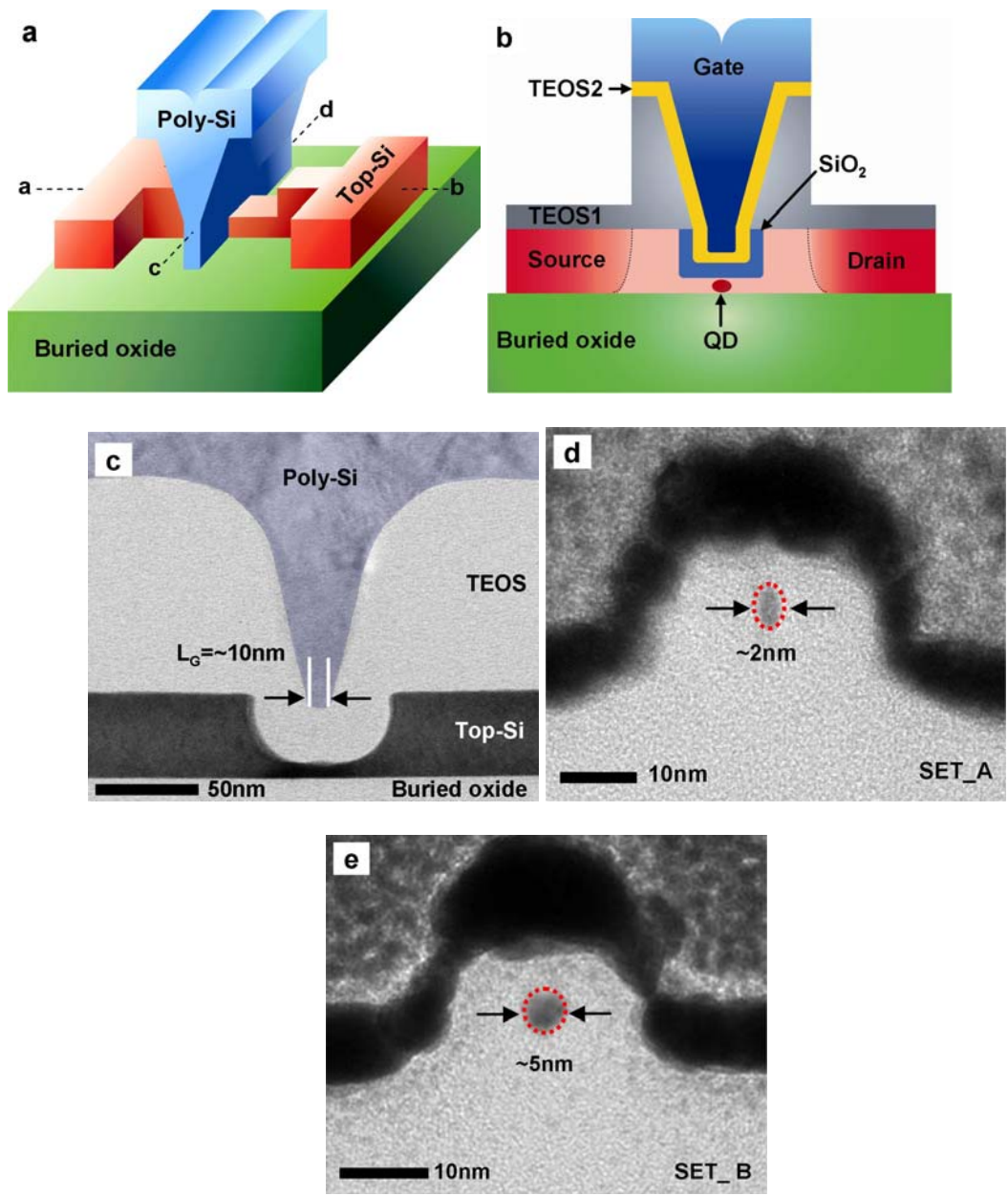

**Figure 1**: Si-SET device schematic diagram and TEM image. a, Schematic 3-D layout and b, cross-sectional view along the channel. c, cross-sectional TEM images along the cutlines a-b and c-d, respectively. d&e, Cross-sectional TEM images of the etched Si wires after oxidation at 900℃ for 50min and 40min, showing Coulomb island sizes of ~2nm for SET_A and ~5nm for SET_B).

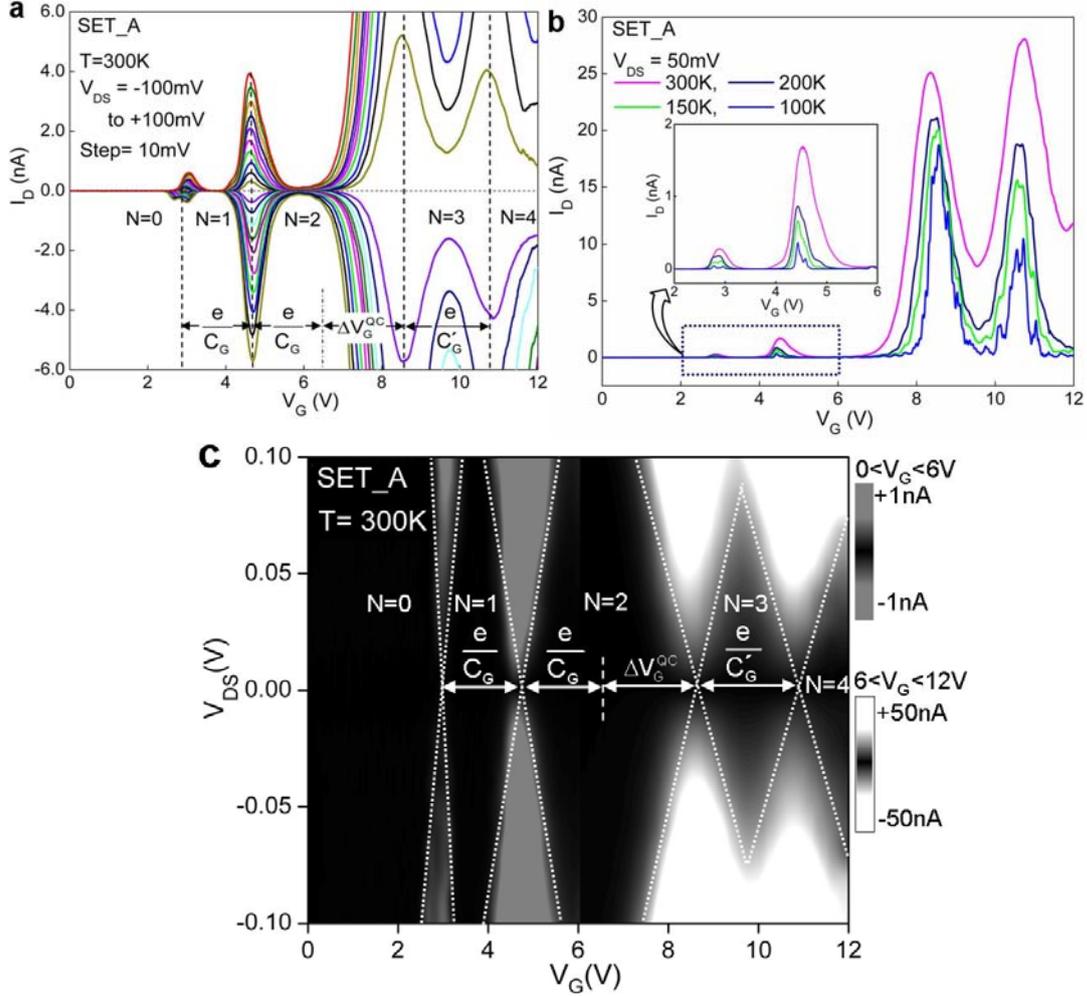

**Figure 2:** Transport characteristics for SET_A. a, drain current measured at 300K as a function of fin-gate voltage $V_G$ using drain voltages from -100 to 100mV with a step of 10mV. The electron occupancies of the island are assigned to the $V_G$ regions between neighboring Coulomb peaks. The Coulomb peak spacing $\Delta V_G^{CB}$ due to Coulomb blockade is denoted by $e/C_G$, while $\Delta V_G^{QC}$ represents the additional gate voltage shift of the 3$^{rd}$ Coulomb peak due to the QC effect. b, Temperature dependence at a fixed drain voltage of $V_{DS}$ =50mV. As temperature decreases from RT to 100K, the Coulomb peak positions are barely changed, but their PVCRs increase. The inset shows detail from the first two Coulomb peaks in the current regime marked with dotted line. c, Stability plot at room temperature; each diamond corresponds to a stable charge configuration state with fixed electron occupancy N.

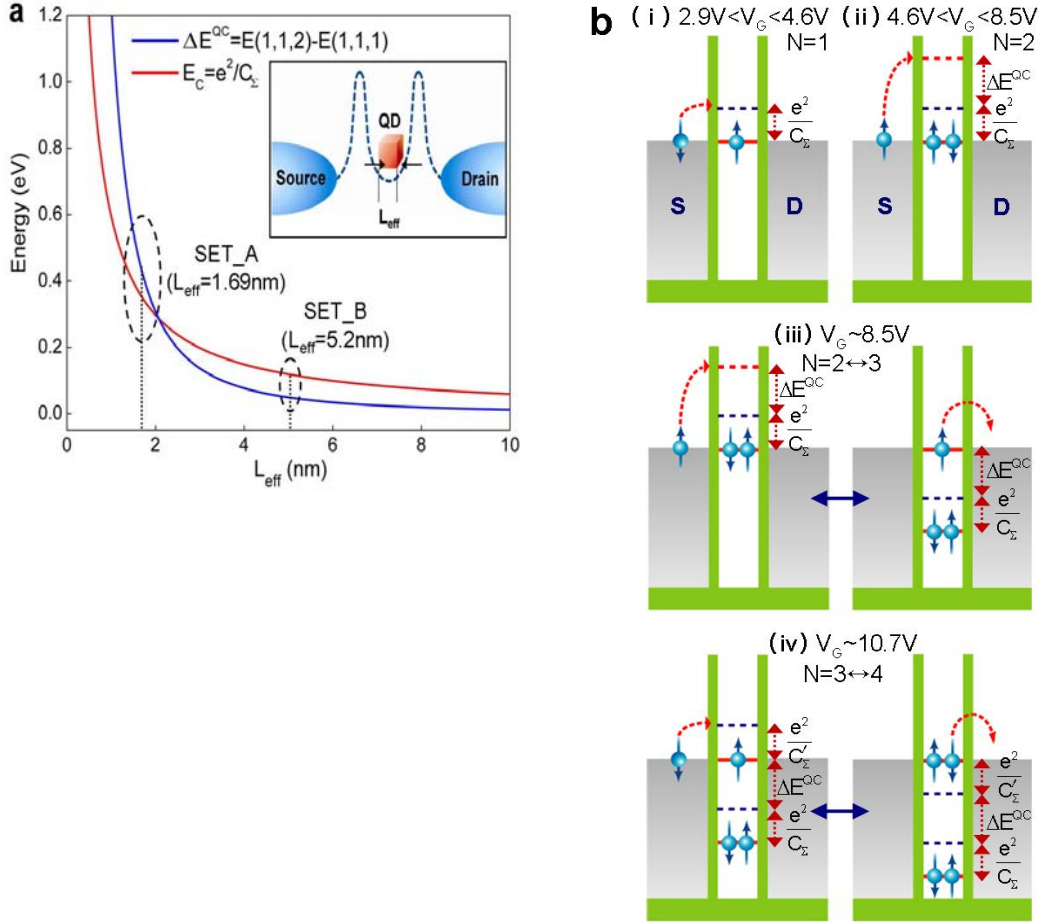

**Figure 3:** Electron transport via an ultra-small Si island. a, Calculations of the Coulomb charging energy $e^2/C_\Sigma$ and a level spacing $\Delta E^{QC}$ due to quantum confinement as a function of the effective side $L_{eff}$ of an ultra-small cubic silicon island. Inset SET schematic consisting of the Si Coulomb island with two identical tunneling barriers connected to source and drain leads. The values of $\Delta E^{QC}$ and $e^2/C_\Sigma$ for the SET_A and SET_B are marked with dotted lines, respectively. b, Schematic energy level diagrams. Each diagram corresponds to a particular gate voltage region seen in Fig. 2c (SET_A); (i) the island occupancy is N=1 for 2.9V<$V_G$<4.6V, corresponding to the 1st Coulomb diamond region. (ii) A two-electron charge configuration state is stable for gate voltages in the range 4.6V<$V_G$<8.5V, corresponding to the 2nd large Coulomb diamond. The diagram shows that the addition of another electron requires a tunneling process into the 1st excited state due to the influence of the Pauli spin exclusion principle, so requiring an energy of $\Delta E^{QC}$ in addition to the Coulomb energy $e^2/C_\Sigma$. (iii) Transport behavior expected around the 3rd Coulomb peak, showing single-electron sequential tunneling via the island. The occupancy of the dot oscillates between N=2 and N=3. (iv) Transport behavior expected around the 4th Coulomb peak, where the dot occupancy oscillates between N=3 and N=4. The figure shows that the addition of another electron requires needs only the Coulomb charging energy.

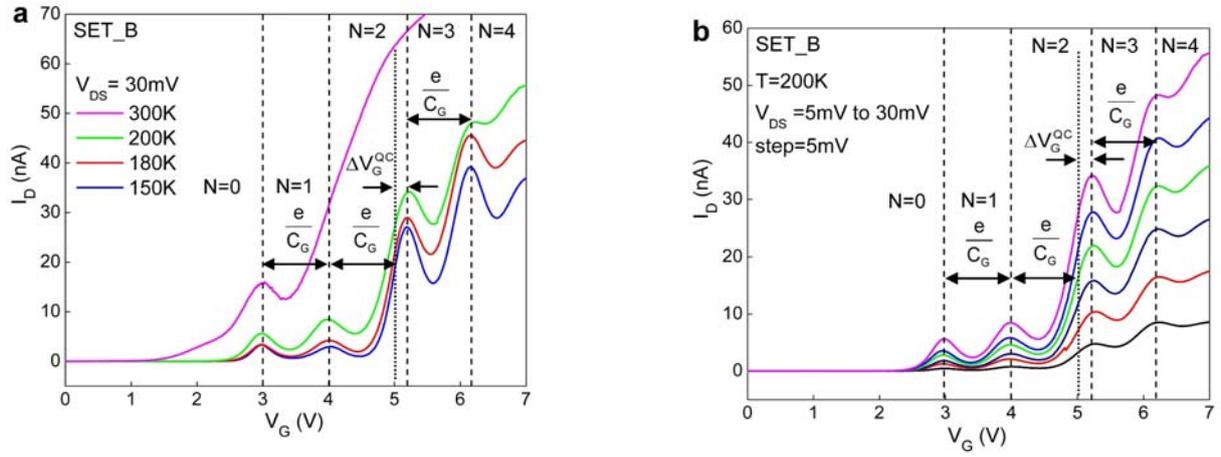

**Figure 4:** Transport characteristics of the SET_B. a, Temperature dependence for a drain bias voltage of $V_{DS}$=30mV. The electron occupancies of the island are assigned to the $V_G$ regions between neighboring Coulomb peaks in a similar manner to those for SET_A. The Coulomb peak spacing $\Delta V_G^{CB}$ due to Coulomb blockade is denoted by $e/C_G$, while $\Delta V_G^{QC}$ represents the additional gate voltage shift of the 3$^{rd}$ Coulomb peak due to the QC effect. b, Device behavior measured at 200K.